\documentclass[showpacs,preprintnumbers,amsmath,amssymb,superscriptaddress]{revtex4}
\usepackage{graphicx}
\usepackage{dcolumn}
\usepackage{bm}
\usepackage{color}
\graphicspath{{EPS_figs/}{PDF_figs/}}

\begin{document}
\title{The ground state and the character of the interaction between a colloidal particles in a liquid crystals}

\author{B. I. Lev}
\affiliation{Bogolyubov Institute for Theoretical Physics, NAS of Ukraine, Metrologichna 14-b, Kyiv 03680,Ukraine}
\affiliation{Nanosystem Research Institute, National Institute of Advanced Industrial Science and Technology (AIST), 1-1-1 Umezono, Tsukuba 305-8568, Japan}

\date{\today}


\begin{abstract}
In this article is proposed the general approach to determination the character of the interaction between colloidal particles in 
a different liquid crystals. The main idea of this approach are in the presentation of the colloidal particle as sours of the possible 
deformation of the ground state of the director field. The ground state of liquid crystal imposes restrictions on the possible deformations 
and as result determine the character of the interaction between the colloidal particles. Based on this approach the Coulomb like 
interaction between dipole particles in a cholesteric liquid crystal and changing of the character of the interaction in a smactic liquid 
crystal has been predicted.
\end{abstract}
\maketitle

Colloidal suspensions in liquid-crystal hosts constitute a new class of composite materials with unique physical properties \cite{b1} 
which are very different from usual colloids \cite{b5}. Particles, suspended in a liquid crystal host, produce the director field 
distortions, which, in turn, give rise to interparticle interactions that do not occur in usual colloids \cite{b6}. These long-range 
anisotropic interactions in liquid crystals lead to formation of different structures such as linear \cite{Poulin_1998, Poulin_1997} 
and inclined \cite{Smalyukh_2005, Kuzmin_2005, Kotar_2006} chains. Particles at a nematic-air interface as well as quasi two-dimensional 
colloids in thin nematic cells form a rich variety of 2D crystals \cite{Nazarenko_2001, Smalyukh_2004, Musevic_2006}. Recently, authors of 
\cite{nych1} observed 3D crystal structures. The understanding of the elasticity-mediated interactions is quite important to predict and 
control properties and behavior of materials based on liquid crystal colloids. Such interactions is of great importance also in various 
processes in more complex media with orientation order, such as solutions of DNA and other biologically relevant molecules.  

Colloidal particles break the continuous symmetry the distribution of the director field which may be accompanied by topological defects 
\cite{Poulin_1998, Stark_1999, Lubensky_1998}. Despite such a complex director field distribution every particle can be effectively presented 
as a point-like source of the deformations of monopole, dipole and quadrupole type \cite{Lubensky_1998}-\cite{fukuda3} with formal analogy 
with classical electrostatics. This fact became a starting point for a number of approaches toward the theory of nematic liquid crystal colloids 
\cite{Lubensky_1998},\cite{Lev_1999}. This paper focused on character of the interaction energy between particles immersed in the different 
liquid crystals. To describe the interaction between the colloidal particles can apply the method developed in \cite{Lev_1999, we, lev3} for
nematic hosts and unified with general symmetry approach to determination of the possible deformation in different liquid crystals. This claim 
imposes restrictions on the possible deformation of initial state of a liquid crystals which contribute to the interaction of the particles. 

In presented concept the interaction are result of the summation of the deformation of elastic director field which produce the surface of 
every particle for different boundary conditions. The character of such interaction directly depend from breaking symmetry in the distribution 
of the director field around particles. The possible deformation of the director field depend from shape of the particles, anchoring energy and 
boundary condition at its surface. Presence or absent the topological singularity around the particle depend only from the anchoring strength. 
The surface of particles break the initial symmetry of the director field. This symmetry of the deformation are main motive for the character 
of the interaction at the far distance \cite{Lev_1999},\cite{lev3}. Interact the particle with as the weak as the strong anchoring conditions, 
but in first case this interaction are smallest as in case the strong anchoring. The interact not the immersed particle, interact the deformation 
area which produce the surface of the particle. The particle are grain or the source of the deformation and the character of the interaction at 
the far distance depend only from the symmetry of this produced deformation \cite{lev3}. As example in article \cite{Naz} was obtained the 
interaction between a colloidal particle and a focused laser beam in a nematic liquid crystal. In the article \cite{Pir} was presented the 
experimental study of the nematic force of the interaction of a colloidal particle with a disclination line. Was show that the electrostatic 
analogy at large distances is basically correct but incomplete since the elastic potential of a disclination line exhibits a lower symmetry 
than its electrostatic analogue. The using the electrostatic analogy help only to understanding the possible form of the interaction energy 
but is not valid for general case. The main idea are in the presentation of interaction between particles immersed in different liquid crystal 
as interaction between possible deformation of ground state, which produced the surface of the particles. 

Our consideration can start from the bulk elastic free energy for different liquid crystals. The elastic bulk free energy of nematic and smectic liquid 
crystals within the one constant approximation can present in the following form
\begin{equation}\label{F_bulk}
F_{\text{def}} = \frac{K}{2} \int dV \left[ (\nabla \cdot \textbf{n} )^2 + (\nabla \times \textbf{n} )^2 \right], 
\end{equation}
where $K$ is the Frank elastic constant. In the case cholesteric liquid crystal should be supplemented an additional term,
\begin{equation}\label{F_ch}
F_{\text{ch}} = K q\int dV (\textbf{n}\cdot \nabla \times \textbf{n}) 
\end{equation}
favoring molecular twist, where $q=\frac{2\pi }{p}$ is wave number which determined through the pitch $p$ of twisted structure of the ground state. 
The cholesteric elastic free energy for long-range deformation are same as in the case smectic liquid crystal with the layered structures, where the 
layer presented the thickness of pitch \cite{de_Gennes}. 

In all liquid crystal the ground state have different symmetry but have same description. Can introduce the general axis along this natural structure.
In a nematic liquid crystal it is the optic axis, in a smectic liquid crystal it is the axis perpendicular to the layers and in a cholesteric liquid 
crystal it is the axis of the pitch. This natural axis we can note as OZ axis. The other direction are perpendicular to this axis and note as $\perp $ 
direction. In all case the ground state the nonuniform director field  can present as $n^{0} (\textbf{r})= \left({\bf n}^{0}_{\perp }, n^{0}_{z} \right)$ 
and exist only one component $n^{0} (\textbf{r})= \left(0,1 \right)$ for nematic and smectic liquid crystal and $n^{0} (\textbf{r}) = \left(1,0 \right)$ 
for the cholesteric liquid crystal, but in last case ${\bf n}^{0}_{\perp }=\left(n^{0}_{x}, n^{0}_{y} \right)= \left(cos qz, sin qz \right)$
in usual coordinate system. 

The deformations which produce a single immersed particle arise as break the symmetry of such ground state. At the short distance the deformation of 
ground state are impossible obtained, because the elastic matter are nonlinearity after the natural condition $\textbf{n}^2(\textbf{r}) = 1$. The topology 
of distribution of the director field needer the particle depend from the strength of anchoring energy. The deviation from the symmetry of the ground state 
can be determine at the far distance if introduce the concept of deformation coat around the particle \cite{lev3}. In this concept the deformation coat 
include all possible nonlinearity and defect structure around the particle. At the far distance we can determine all possible small deformation of the 
director field which take into account the breaking symmetry of the distribution director field on the short distance from particle \cite{Lev_1999,lev3}. 
In this approach can use only small deformation of director field which depend from the geometrical shape of particle or the deformation coat. 

The every immersed particle by the shape and the boundary condition on the surface produce the deformation of the ground state ${\bf n}_{0}$ which are small 
everywhere at the far distance : ${\bf n}({\bf r})={\bf n}_{0}+\delta {\bf n}({\bf r}),\left| \delta {\bf n}\right| \ll 1$ and ${\bf n}_{0}\cdot \delta {\bf n}({\bf r})=0$. 
Can introduce the same component of the deformation of the director field as follows $ \delta {\bf n}({\bf r})=\left(\delta {\bf n}_{\perp },\delta n_{z } \right)$.  
In the case of nematic and smectic liquid crystals we have only $ \delta {\bf n}({\bf r})=\left(\delta {\bf n}_{\perp },0 \right)$
and in the case cholesteric liquid crystal we have components $ \delta {\bf n}({\bf r})=\left(\delta {\bf n}_{\perp },\delta n_{z } \right)$. For description 
deformation in different liquid crystal was introduce the different additional variables. In the case nematic liquid crystal $ \delta {\bf n}_{\perp }({\bf r})=\left(\delta n_{x},\delta n_{y } \right)$
in the case smectic liquid crystal $\delta {\bf n}_{\perp }=-\nabla_{\perp } u,$ where $ u ({\bf r})$ is displacement of the layer and  i.e., 
$\ \delta n_{x}=-\frac{\partial u}{\partial x},\delta n_{y}=-\frac{\partial u}{\partial y}$  \cite{Chaikin}. Definition $\nabla_{\perp }$ is the gradient 
operator in the $x-y$ plane along layer. In the case cholesteric liquid crystal $ \delta {\bf n}({\bf r})=\left(-u\sin qz, u\cos qz,v\right)$ where $u({\bf r})$ 
and $v({\bf r})$ describe the umbrella and azimuthal deformation of ground state. 

In this notation of the deformation of the ground state can present the bulk free energy in the case nematic liquid crystal in the simple form 
\begin{equation}
F_{N \text{bulk}} = \frac{K}{2} \int dV  (\nabla \delta n_{\perp })^2  
\end{equation}
Phenomenologically the free energy of a smectic liquid crystal within the continuous approximation can be written in the following harmonic representation \cite{Chaikin}
\begin{equation}
F_{S \text{bulk}}=\frac{1}{2}\int d\textbf{r} \left[B(\nabla_{z}u)^{2}+K(\nabla_{\perp}^{2}u)^{2}\right], 
\end{equation}
where $B$ is a compression modulus and $K$ is the splay elastic constant. Between compression modulus and elastic constant can introduce the relation $B=\frac{K}{\lambda^{2}}$ 
where $\lambda$ is characteristic length, typically of order of the layer spacing. The second part this free energy came from Frank splay elastic energy which same as nematic
liquid crystal and first part take into account only possible compression layers. The nonlinear part of the free energy, neglected in bulk free energy for small deformation \cite{Chaikin}. 
In the terms of small $u(\textbf{r})$ and $v(\textbf{r})$ the bulk free energy of cholesteric liquid crystal can be present as:
\begin{equation}
F_{Ch \text{bulk}} = \frac{K}{2} \int dV  \left[ (\nabla v)^2 +(\nabla u)^2 +q^2 v^2 
+q\left(\frac{\partial u}{\partial x}v-\frac{\partial v}{\partial x}u\right)\cos qz + q\left(\frac{\partial u}{\partial y}v- \frac{\partial v}{\partial y}u\right)\sin  qz \right], 
\end{equation}
which for $q=0$ completely reproduces the bulk free energy of a nematic liquid crystals \cite{we,we2}. If neglect \cite{we2} the influence of the off-diagonal part 
 which can not change the character of the interaction between particles, the bulk free energy of cholesteric liquid crystal can present in the simples form
\begin{equation}
F_{Ch \text{bulk}} = \frac{K}{2} \int dV  \left[ (\nabla v)^2 +(\nabla u)^2 +q^2 v^2 \right], 
\end{equation}

The possible deformation of the elastic director field produce the surface of every particle. The surface free energy independence from type of liquid crystal and is given by 
the well-known Rapini-Popular form
\begin{equation}
F_{s}=\sum_{p}\oint dsW({\bf s})({\bm{\nu} }({\bf s})\cdot {\bf n}({\bf s}))^{2}.  \label{3}
\end{equation}
where $ \bm{\nu}$ is the unit normal at the point $\mathbf{s}$ on the surface of the $p$-th particle. In the case of homeotropic anchoring, the anchoring strength $W(s)$ is 
negative and is positive for planar anchoring. Due to the anchoring on its surface every particle gives rise to the distortions of the liquid crystal ordering. These distortions, 
in turn, depend on the particle shape as well as on the anchoring strength. We restrict ourselves to the case of weak anchoring $\frac{W R}{K} \ll 1$, where $R$ is the 
characteristic size of the particle. In the case of strong anchoring can use the same representation of a source of the deformations if take into account the director 
distribution around a single particle \cite{we,we2,fukuda2,fukuda}. Therefore, hereinafter we suppose that the director deviations from its ground state ${\bf n}_{0}$ 
are small everywhere. Under these conditions the director field $\mathbf{n}(\textbf{r})$ is defined and continuous through out the system, even within the particles. 
Then the relevant terms in $F_s$ are first order in $\delta\mathbf{n}$ and can rewrite in the form
\begin{equation}\label{F_s_series}
F_{s} \simeq  2\sum_{p}\oint d \mathbf{s} W(\mathbf{s})\{\bm{\nu}(\mathbf{s})\cdot\mathbf{n}_0(\mathbf{s})\}
\{\bm{\nu}(\mathbf{s})\cdot\delta\mathbf{n}(\mathbf{s})\}=2\sum_{p}\oint d \mathbf{s} W(\mathbf{s})
\{\bm{\nu}(\mathbf{s})\cdot \widehat{L}(\mathbf{s})\mathbf{n}_0(\mathbf{r_p}\}\{\bm{\nu}(\mathbf{s})\cdot \widehat{L}(\mathbf{s}) \delta\mathbf{n}(\mathbf{r_p})\}
\end{equation}
where operator $\widehat{L}(\mathbf{s})=[1 + (\bm{\rho}\cdot\nabla)+(1/2) (\bm{\rho}\cdot\nabla)(\bm{\rho}\cdot\nabla)]$ determine slowly spatial 
dependence as ground state as the deformation of director field, $\bm{\rho}$ present the radius vector to surface point from spatial point of particle 
position $\mathbf{r_p}$ \cite{Lev_1999, lev3}. In the case of nematic and smectic liquid crystals the $\mathbf{n}_0(\mathbf{s})$ are not depend from spatial position of 
the particle and we can take the director field in nonuniform form, but in the case cholesteric liquid crystal the ground state is depend from spatial position and only 
in this case we should be take into account the presented expression for determination director field on the surface of the single particle. In all case can represent 
$F_s$ in the following compact form 
\begin{equation}
F_{s}=\sum_{p}\widehat{A}_p^{i}\delta {\bf n}_{i}(\mathbf{r}_p)
\end{equation}
where operator $\widehat{A}_p^{i}$ is  determined by the particle shape \cite{Lev_1999,lev3} and the director ground state 
$ \widehat{A}^{i}_{p} = \oint d \bm{s} W(\bm{s})(\bm{\nu}(\mathbf{s})\cdot \widehat{L}(\mathbf{s})\mathbf{n}_0(\mathbf{r_p}) \bm{\nu}_{i} \widehat{L}(\bm{s})$.	
Such a treatment of the surface energy represents a powerful technique as it allows us to extract the far-field interaction the colloidal particle 
of the arbitrary shape.

The distortion profile that minimizes the elastic energy in the presence of the particles can be determined from the following Euler-Lagrange equations
$\frac{\delta \left( F_{\text{bulk}}+F_{s}\right) }{\delta (\delta \bm{n}_{\perp })}=\frac{\delta \left( F_{\text{bulk}}+F_{s}\right) }{\delta
(\delta \bm{n}_{z})} =0$. Solutions to these equations can be written via appropriate the Green functions $ \delta \bm{n}_{\perp }= -\frac{1}{4 \pi K}\sum_{p}\int d {\bf r}^{\prime } \widehat{A}^{\perp }_{p}\delta({\bf r}-{\bf r}_{p}^{\prime })G_{\perp }({\bf r}-{\bf r}^{\prime })$
$ \delta \bm{n}_{z }= -\frac{1}{4 \pi K}\sum_{p}\int d {\bf r}^{\prime } \widehat{A}^{z}_{p}\delta({\bf r}-{\bf r}_{p}^{\prime })G_{z}({\bf r}-{\bf r}^{\prime })$
where $G_{\perp }$ and $G_{z}$ is Green function the Euler-Lagrange equation for respective deformation director field. Due to the linearity of the Euler-Lagrange equations 
we can use the superposition principle for the system of many colloidal particles. That is, the director field distortions are the sum of distortions caused by every single 
particle. Substituting solutions  into the free energy and implying the superposition principle we come to the fact that $F_{\text{bulk}}+F_{s} = \sum_{p>p^{\prime}}U_{p,p^{\prime}}+\sum_{p}U_{p}$, 
where $U_{p}$ is the self-energy of the $p$-th particle and $U_{p,p^{\prime}}$ is the energy of the interaction between the $p$-th and $p^{\prime}$-th particles \cite{Lev_1999, lev3,we,we2}
\begin{equation}
U_{p,p^{\prime }}=-\frac{1}{4 \pi K} \left[ \widehat{A}^{\perp }_{p}\widehat{A}^{\perp }_{p^{\prime}} G_{\perp }({\bf r}-{\bf r}^{\prime })+ \widehat{A}^{z}_{p}\widehat{A}^{z }_{p^{\prime}} G_{z}({\bf r}-{\bf r}^{\prime })\right]
\end{equation}
This expression is the most general representation of the pair interaction energy in different liquid crystal colloids. Next can consider of the particles 
interaction in different liquid crystal. 

In the case nematic liquid crystal we have only one component of deformation director field $\delta \bm{n}_{\perp }$. The surface energy take the form 
$F_{s}=\sum_{p}A^p_{\perp } \delta \bm{n}_{\perp }(\bm{r_p})$ where $A^p_{\perp }=2 \oint d \mathbf{s} W(\mathbf{s})\bm{\nu}_z \bm{\nu}_{\perp } \widehat{L}(\bm{s})= 
2 \oint d \mathbf{s} W(\mathbf{s})\bm{\nu}_z \bm{\nu}_{\perp } (1 + (\bm{\rho}\cdot\nabla)+(1/2) (\bm{\rho}\cdot\nabla)(\bm{\rho}\cdot\nabla))$  which can be presented
in the form multiple expansion $A^p_{\perp }= (q_{z,\perp } + p_{z,z,\perp }\cdot\nabla{z}+(1/2) Q_{z,z,z,\perp }\nabla^2{z})$. Every introducing multiple coefficient
presented trough the integral over surface of particle : $q_{z,\perp }=2 \oint d \mathbf{s} W(\mathbf{s})\bm{\nu}_z \bm{\nu}_{\perp }$ , 
$p_{z,z,\perp }=2 \oint d \mathbf{s} W(\mathbf{s})\bm{\nu}_z \bm{\nu}_{\perp } \bm{\rho}_{z }$ and $Q_{z,z,z,\perp }= \oint d \mathbf{s} W(\mathbf{s})\bm{\nu}_z \bm{\nu}_{\perp }\bm{\rho}_{z } \bm{\rho}_{z }$
and independence from any other condition, only from its shape. If take into account the Euler-Lagrange equation in the form $\Delta n_{\perp } = \sum_{p}A^p_{\perp }$,
as well as in classical electrostatics, solutions of these equations can be expanded in presented multiples $ n_{\perp } (\textbf{r}) = \frac{q_{z,\perp }}{r} + \frac{p_{z,z,\perp }}{r^3} + \frac{Q_{z,z,z,\perp }}{r^5} +...$
which is valid at large distances. Hence, the long-range interactions in NLC colloids are always controlled by multiple coefficients. This presentation take place 
that the Euler-Lagrange equation have the same form as equation for electrostatic potential. From this presentation can obtain the interaction energy for particles. 
The considered ground state of nematic liquid crystal let make the next conclusion, that the Coulomb like interaction can be exist if the surface breaking the all 
elements of symmetry of continuous distribution director field  \cite{lev3}. The dipole moment appear if break the horizontal elements of symmetry and quadrupole 
moment exist for the presence as horizontal as vertical elements of symmetry on the deformation area which produce the surface of the particles or the deformation 
coat  \cite{lev3}. We have only one dipole moment, which can be presented as two component which related to usual coordinate system. The real dipole moment are parallel 
to director field of the ground state.

In the case of smectic liquid crystal in the term introducing new variable which determine the displacement of the layer the surface energy in general it can be done in 
the following form $ F_{s}=\sum_{p}\widehat{A}_{p}({\bf r})u({\bf r}_p)$ where operator $\widehat{A}_{p}({\bf r})$ is determined by the shape of the $p$-th particle. 
In the case of particles with arbitrary shape and weak anchoring $F_{s}=-\sum_{p}(q_{z,\perp } + p_{z,z,\perp }\cdot\nabla{z}+(1/2) Q_{z,z,z,\perp }\nabla^2{z}) \nabla_{\perp }u$ 
\cite{we,we2,fukuda2,fukuda}. where deformation charge $q_{z,\perp }$  dipole $p_{z,z,\perp }$ and quadrupole $Q_{z,z,z,\perp }\nabla^2{z}$ have the same presentation 
as in the case of nematic liquid crystal. From this presentation we can conclude that the particle which produce the deformation charge will be interact as dipole 
particle and particle which have the dipole moment have only quadrupole interaction between such particle. For asymmetric particle the deformation charge absent and 
first multiple expansion will be start from dipole-dipole interaction, but in smectic liquid crystal this interaction have the quadrupole type.

The distortion profile that minimizes the elastic energy in the presence of the particles can be determined from the Euler-Lagrange equation  can determine 
Fourier presentation of displacement $ u({\bf q})=-\frac{2}{B}\sum_{p}\frac{A^{*}_{p}({\bf q})}{\left[q^{2}_{z}+\lambda^{2}q^{4}_{\perp}\right]}$
where $\widehat{A}_{p}({\bf q})$ is the Fourier representation of operator $\widehat{A}_{p}({\bf r})$. Substituting the solution into the free energy of 
the system one can easily find the interaction energy between $p$-th and $p^{\prime}$-th particles in the form 
$U_{pp^{\prime }}^{0} =\frac{1}{\pi^{3}B}\widehat{A}_{p}\widehat{A}_{p^{\prime } } \partial_{\perp}\partial_{\perp}^{\prime}\int d {\bf q} \frac{e^{-i{\bf qr}}}{
q_{z }^{2}+\lambda ^{2}q_{\perp }^{4}}$, where ${\bf r}=$ ${\bf r}_{p} - {\bf r}_{p^{\prime}}$ denotes the interparticle distance. From this presentation can conclude 
that the interaction energy trough the introducing moment of electrostatic analogy but the Green function in this case have the other presentation and 
not related with this analogy.

We can note about the principal moment of motivation general character of interaction between particles in different liquid crystal with presented 
structure of ground state of the distribution of elastic director field. In this case cholesteric liquid crystal the Green functions for both introducing 
variable have the analogy with electrostatic, usual Coulomb Green function for $u(\textbf{r})$ variable $G_{u}(\textbf{r}, \textbf{r}^{\prime}) = \frac{1}{|\textbf{r} - \textbf{r}^{\prime}|}$,
and screening Coulomb Green function $G_v(\textbf{r}, \textbf{r}^{\prime})=  \frac{1}{|\textbf{r} - \textbf{r}^{\prime}|}exp (q|\textbf{r} - \textbf{r}^{\prime}|)$ 
for variable $v(\textbf{r})$. But in this case we have problem with interpretation the source free energy, which can be presented only in the combination 
of introducing constant as deformation charge, dipole and quadrupole moment. In the case of cholesteric liquid crystal this moments are not constant and 
depended from spatial point. In the case of bigger particles, when take into account the distribution director field on the surface of every particle should
be consider the next presentation of the surface free energy 
$F_{s}=2\sum_{p}\oint d \mathbf{s} W(\mathbf{s})\bm{\nu}_{\perp } \bm{n}^{0}_{\perp }(\bm{r_p}) \{ \bm{\nu}_{\perp } \widehat{L}(\bm{s}) \delta \bm{n}_{\perp }(\bm{r_p})+\bm{\nu}_z \widehat{L}(\mathbf{s}) \delta \bm{n}_{z}(\bm{r_p})\}$
from which can see that in the cholesteric liquid crystal can exist two deformation charge and exist the Coulomb like interaction for bigger particle 
with dipolar configuration of the ground state. For small particle this effect absent.

Two particles suspended in a liquid crystals attract or repel each other due to an overlapping of the director field deformations they produce.
Within the continuous description of liquid crystal ordering only the distortions which produce the symmetry of ground state of distribution of 
the director field can exist. Using this suggestion we derived the interaction potential valid for colloidal particles of different shapes. In 
presented approach can determine a character of the interaction energy for different symmetry of ground state. The range of the potential matches 
with the decay length of the elastic deformations and the interaction is sufficient for the formation of various structures, which can be observed 
experimentally. The effective pair potentials based on the linear theory of elasticity are restricted to the case of small deviations from a ground 
state of the liquid crystal but have the general presentation and prediction. 

Author is grateful to the Japan Society for the Promotion of Science (JSPS) for the financial support to stay at National Institute of Advanced
Industrial Science and Technology (AIST) and to carry out this work there.

\end{document}